\documentstyle[pra,aps,epsf]{revtex}

\newcommand{\bfc}{{\bf c}}
\newcommand{\bfK}{{\bf K}}
\newcommand{\bfL}{{\bf L}}
\newcommand{\bfp}{{\bf p}}
\newcommand{\bfP}{{\bf P}}
\newcommand{\bfQ}{{\bf Q}}
\newcommand{\bfR}{{\bf R}}
\newcommand{\bfu}{{\bf u}}
\newcommand{\bfv}{{\bf v}}
\newcommand{\bfx}{{\bf x}}

\newcommand{\bfnabla}{{\mbox{\boldmath $\nabla$}}}

\newcommand{\collf}{{\Omega}}

\newcommand{\bfone}{{\bf 1}}
\newcommand{\calDc}[2]{{\cal D}^{#1}_{#2}}
\newcommand{\ene}{{\varepsilon}}
\newcommand{\Ene}{{\cal E}}
\newcommand{\dt}{{\Delta t}}
\newcommand{\prop}{\left(\bfx+\epsilon\bfc_i,t+\epsilon\dt\right)}
\newcommand{\here}{\left(\bfx,t\right)}
\newcommand{\Dopi}[1]{{\hat{D}_{#1}}}
\newcommand{\Dopil}[2]{{\hat{D}_{#1}^{({#2})}}}
\newcommand{\pt}[1]{{\partial_{t_{#1}}}}
\newcommand{\ptp}[2]{{\partial_{t_{#1}}^{#2}}}
\newcommand{\cvop}[2]{{\pt{#2} + \bfv_{#1}\cdot\bfnabla}}
\newcommand{\smallfrac}[2]{{\mbox{\small $\frac{#1}{#2}$}}}

\begin{document}

\title{
  \begin{flushleft}
    {\footnotesize BU-CCS-980602}\\[1.0cm]
  \end{flushleft}
  \bf Inverse Chapman-Enskog Derivation of the Thermohydrodynamic
  Lattice-BGK Model for the Ideal Gas~\footnote{To appear in {\it
  International Journal of Modern Physics C} (1998)}}
\author{
  Bruce M. Boghosian\\
  {\footnotesize Center for Computational Science, Boston University,}\\
  {\footnotesize 3 Cummington Street, Boston, Massachusetts 02215, U.S.A.}\\
  {\footnotesize{\tt bruceb@bu.edu}}\\[0.3cm]
  Peter V. Coveney\\
  {\footnotesize Schlumberger Cambridge Research,}\\
  {\footnotesize High Cross, Madingley Road, Cambridge CB3 OE5, U.K.}
  }
\date{\today}
\maketitle

\begin{abstract}
A thermohydrodynamic lattice-BGK model for the ideal gas was derived by
Alexander et al.\ in 1993, and generalized by McNamara et al.\ in the
same year.  In these works, particular forms for the equilibrium
distribution function and the transport coefficients were posited and
shown to work, thereby establishing the {\it sufficiency} of the model.
In this paper, we rederive the model from a minimal set of assumptions,
and thereby show that the forms assumed for the shear and bulk
viscosities are also {\it necessary}, but that the form assumed for the
thermal conductivity is not.  We derive the most general form allowable
for the thermal conductivity, and the concomitant generalization of the
equilibrium distribution.  In this way, we show that it is possible to
achieve variable (albeit density-dependent) Prandtl number even within a
single-relaxation-time lattice-BGK model.  We accomplish this by
demanding analyticity of the third moments and traces of the fourth
moments of the equilibrium distribution function.  The method of
derivation demonstrates that certain undesirable features of the model
-- such as the unphysical dependence of the viscosity coefficients on
temperature -- cannot be corrected within the scope of lattice-BGK
models with constant relaxation time.
\end{abstract}

\vspace{0.2truein}

\section{Introduction}

A thermohydrodynamic lattice BGK model for the ideal gas was derived by
Alexander et al.\ in 1993~\cite{bib:alexander}.  In the same year,
McNamara and coworkers~\cite{bib:mac} generalized this model to include
different relaxation times for the stress and the energy, thereby
allowing adjustment of the Prandtl number.  These analyses assumed
particular forms for the equilibrium distribution function, took the
transport coefficients to be linear in the internal energy $I$ and
independent of the density $\rho$, and carried out a Chapman-Enskog
analysis to show that these choices yielded the desired
thermohydrodynamic equations.  In this way, the {\it sufficiency} of the
thermohydrodynamic lattice-BGK model was established.  In this paper, we
rederive this model from the demand that the moments of the equilibrium
distribution be analytic.  We show that the forms assumed for the shear
and bulk viscosities in earlier works are {\it necessary} as well as
sufficient, but that the form assumed for the thermal conductivity $k$
is not.  We derive the most general possible form for $k$, as well as
the alteration that this incurs in the form of the equilibrium
distribution function, and we show that this enables some adjustment of
Prandtl number even within a single-relaxation-time lattice-BGK model.

We demonstrate this by carrying out an ``inverse Chapman-Enskog
analysis''; that is, we assume only that the first three moments and the
trace of the fourth moment of the equilibrium distribution function
exist and are analytic, and we demand that the resulting macroscopic
equations be identical to the usual ones of thermohydrodynamics.  We
then work backwards to find the requirements imposed upon the
equilibrium and the transport coefficients.  As shall be shown, this
process determines the derivatives of the third moments and the traces
of the fourth moments of the equilibrium distribution function with
respect to the conserved densities.  Analyticity of the moments then
requires that their mixed second derivatives be symmetric with respect
to the order of differentiation.

It is sometimes glibly stated that this ``inverse Chapman-Enskog''
approach to deriving lattice Boltzmann models is capable of dealing with
essentially any hydrodynamic equations whatsoever.  This work shows that
this is definitely {\it not} the case.  In fact, as we shall see, even
the problem for the ideal gas is hugely overdetermined.  Such
overdetermination is a serious problem with lattice BGK models in spite
of the freedom that one has in choosing different lattices and different
sets of particle velocities.  Indeed, within the context of a lattice
BGK model for the usual equations of viscous thermohydrodynamics, with
constant relaxation parameter, this paper will demonstrate that {\it no}
choice of lattice or of velocity set will yield viscosities that vary as
the square root of temperature, as is usually desired.  This follows
from nothing more than the requirement that the moments of the
equilibrium distribution be analytic.  It may be possible to circumvent
this restriction by other generalizations of the BGK methodology -- for
example, by introducing a particular dependence of the relaxation
parameter on the hydrodynamic densities -- but this has yet to be
demonstrated.

\section{The Chapman-Enskog Analysis}

As usual, we assume that the mass, momentum and kinetic energy densities
are given by the first three moments of the discrete-velocity
distribution function,
\begin{eqnarray*}
\rho &=& \sum_i f_i = \sum_i f^{(0)}_i\\
\bfp &=& \sum_i \bfc_i f_i = \sum_i \bfc_i f^{(0)}_i\\
\Ene &=& \sum_i \frac{c_i^2}{2} f_i = \sum_i \frac{c_i^2}{2} f^{(0)}_i,
\end{eqnarray*}
where $f^{(0)}_i$ is the equilibrium distribution.  The kinetic equation
can be written
\begin{equation}
f_i\prop = f_i\here + \collf_i,
\label{eq:kin0}
\end{equation}
where we have denoted the lattice vectors by $\bfc_i$, and the time step
by $\dt$ which is equal to unity in natural lattice units.  We have also
introduced the gradient expansion parameter $\epsilon$ which will aid in
the Chapman-Enskog analysis, and the collision operator $\collf_i$ which
is required to conserve mass, momentum and kinetic energy
\begin{eqnarray*}
\sum_i\collf_i &=& 0\\
\sum_i\bfc_i\collf_i &=& 0\\
\sum_i\frac{c_i^2}{2}\collf_i &=& 0.
\end{eqnarray*}
The assertion that the energy is purely kinetic will lead us inevitably
to an ideal gas equation of state.

Throughout this work, we assume that the collision operator is of BGK
form,
\begin{equation}
\collf_i = \frac{1}{\tau}\left(f^{(0)}_i - f_i\right),
\label{eq:bgk0}
\end{equation}
where $\tau$ is a characteristic collisional relaxation time.  As noted
above, this model was generalized in subsequent work by McNamara et
al.~\cite{bib:mac} to include more than one relaxation time and thereby
allow for variable Prandtl number.  In this work, we adhere to the
original form with a single relaxation time.  We shall show that some
variation of Prandtl number is possible even in this case.  We
emphasize, however, that there is nothing preventing the application of
our methods to the more general case of multiple relaxation times, and
that such application may uncover still more general lattice-BGK models
for that case as well.

To carry out the Chapman-Enskog analysis, we first solve
Eq.~(\ref{eq:bgk0}) for the full distribution function
\begin{eqnarray*}
f_i
 &=& f^{(0)}_i - \tau\collf_i\\
 &=& f^{(0)}_i - \tau\left[f_i\prop - f_i\here\right]\\
 &=& f^{(0)}_i - \tau\sum_{\ell=1}^\infty
     \frac{\epsilon^\ell}{\ell !}
     \left(\partial_t + \bfv_i\cdot\bfnabla\right)^\ell f_i\\
 &=& f^{(0)}_i - \tau\sum_{\ell=1}^\infty
     \frac{\epsilon^\ell}{\ell !}\left(\Dopi{i}\right)^\ell f_i\\
\end{eqnarray*}
where $\Dopi{i}\equiv\partial_t + \bfv_i\cdot\bfnabla$ and
$\bfv_i\equiv\bfc_i/\dt$, and where Eq.~(\ref{eq:kin0}) was used to get
from the first line to the second.  Next, we establish multiple time
scales by ordering $\Dopi{i}$ as follows:
\[
\Dopi{i} = \sum_{k=1}^\infty \epsilon^{k-1}\Dopil{i}{k},
\]
where
\[
\Dopil{i}{k}\equiv\cvop{i}{k}.
\]
Here $t_1$ is the fastest time scale, $t_2$ is slower, etc.  This
results in
\[
f_i = 
\left[ 1 + \tau
\sum_{\ell=1}^\infty
\frac{\epsilon^\ell}{\ell !}
\left(\sum_{k=1}^\infty\epsilon^{k-1}\Dopil{i}{k}\right)^\ell
\right]^{-1}
f^{(0)}_i,
\]
and we can expand this in $\epsilon$ to obtain
\[
f_i = \sum_{k=0}^\infty \epsilon^k f_i^{(k)},
\]
where to ${\cal O}(\epsilon^2)$,
\begin{eqnarray*}
f_i^{(1)} &=& -\tau\Dopil{i}{1} f_i^{(0)}\\
f_i^{(2)} &=& -\tau\left[\Dopil{i}{1}
              -\left(\tau-\frac{1}{2}\right)
               \left(\Dopil{i}{1}\right)^2\right]f_i^{(0)},
\end{eqnarray*}
or, in expanded form
\begin{eqnarray}
f_i^{(1)} &=& -\tau\left(\cvop{i}{1}\right) f_i^{(0)}
\label{eq:ce10}\\
f_i^{(2)} &=& -\tau\left[\left(\cvop{i}{2}\right)
              -\left(\tau-\frac{1}{2}\right)
               \left(\cvop{i}{1}\right)^2\right]f_i^{(0)}.
\label{eq:ce20}
\end{eqnarray}

\section{The Form of the Equilibrium Distribution}

To proceed, we adopt notation for the first five completely symmetric
tensor moments of the {\it equilibrium} distribution function.  Since
the local equilibrium is entirely determined by the conserved
quantities, all of these moments must be functions of the three
conserved densities, thus
\begin{eqnarray*}
\rho
&=&
\sum_i f_i^{(0)}\\
p_\alpha
&=&
\sum_i c_{i\alpha} f_i^{(0)}\\
Q_{\alpha\beta}(\rho,\bfp,\Ene)
&=&
\sum_i c_{i\alpha} c_{i\beta} f_i^{(0)}\\
R_{\alpha\beta\gamma}(\rho,\bfp,\Ene)
&=&
\sum_i c_{i\alpha} c_{i\beta} c_{i\gamma} f_i^{(0)}\\
S_{\alpha\beta\gamma\delta}(\rho,\bfp,\Ene)
&=&
\sum_i c_{i\alpha} c_{i\beta} c_{i\gamma} c_{i\delta} f_i^{(0)},
\end{eqnarray*}
where Greek letters denote spatial indices.  We also adopt notation for
the traces of these tensors,
\begin{eqnarray*}
Q_{\alpha\alpha}(\rho,\bfp,\Ene)
&=&
2\Ene\\
R_{\alpha\alpha\beta}(\rho,\bfp,\Ene)
&=&
2K_\beta(\rho,\bfp,\Ene)\\
S_{\alpha\alpha\beta\gamma}(\rho,\bfp,\Ene)
&=&
2L_{\beta\gamma}(\rho,\bfp,\Ene).
\end{eqnarray*}

We have not shown and shall not show in this paper how to find a
functional form for an equilibrium distribution with these moments.
Suffice it to say that the need to be able to do this places well
understood constraints on the lattice symmetry group~\cite{bib:lga}.
This is why, for example, a triangular lattice is often used in two
dimensions, and a face-centered cubic lattice is often used in three.
Henceforth, we shall simply assume that the lattice is such that moments
of the distribution transform like isotropic tensors to rank four.

\section{The First-Order Solution}

Taking moments of Eq.~(\ref{eq:ce10}), we get the conservation equations
\begin{eqnarray}
\pt{1}\rho &+& \bfnabla\cdot\bfp = 0\nonumber\\
\pt{1}\bfp &+& \bfnabla\cdot\bfQ = 0\nonumber\\
\pt{1}\Ene &+& \bfnabla\cdot\bfK = 0.\label{eq:fo0}
\end{eqnarray}
These equations tell us how the conserved densities vary on the fastest
($t_1$) time scale.  Since all of the moments are assumed to be
functionally dependent on the conserved densities, we also know how they
vary on the fastest time scale.

\section{The Second-Order Solution}

Taking moments of Eq.~(\ref{eq:ce20}), we get the conservation equations
\begin{eqnarray*}
\pt{2}\rho + \bfnabla\cdot\bfp &=&
\left(\tau-\frac{1}{2}\right)
\bfnabla\cdot
\left[
\ptp{1}{2}\rho +
2\pt{1}\bfnabla\cdot\bfp +
\bfnabla\bfnabla : \bfQ
\right]\\
\pt{2}\bfp + \bfnabla\cdot\bfQ &=&
\left(\tau-\frac{1}{2}\right)
\bfnabla\cdot
\left[
\ptp{1}{2}\bfp +
2\pt{1}\bfnabla\cdot\bfQ +
\bfnabla\bfnabla : \bfR
\right]\\
\pt{2}\Ene + \bfnabla\cdot\bfK &=&
\left(\tau-\frac{1}{2}\right)
\bfnabla\cdot
\left[
\ptp{1}{2}\Ene +
2\pt{1}\bfnabla\cdot\bfK +
\bfnabla\bfnabla : \bfL.
\right]\\
\end{eqnarray*}
Using the first-order solutions, Eqs.~(\ref{eq:fo0}), to simplify the
above, and converting to index notation, we have
\begin{eqnarray}
\pt{2}\rho + \nabla_\beta p_\beta &=& 0\label{eq:som0}\\
\pt{2}p_\alpha + \nabla_\beta Q_{\beta\alpha} &=&
\left(\tau-\frac{1}{2}\right)
\nabla_\beta
\left(
\pt{1}Q_{\beta\alpha} + \nabla_\gamma R_{\gamma\beta\alpha}
\right)\label{eq:sop0}\\
\pt{2}\Ene + \nabla_\beta K_\beta &=&
\left(\tau-\frac{1}{2}\right)
\nabla_\beta
\left(
\pt{1}K_\beta + \nabla_\gamma L_{\gamma\beta}
\right)\label{eq:soe0}.
\end{eqnarray}

To proceed, we make use of the assumed functional dependence of $\bfQ$
and $\bfK$ on the conserved densities, $\rho$, $\bfp$ and $\Ene$.  Using
the chain rule and Eqs.~(\ref{eq:fo0}), we have
\begin{eqnarray*}
\pt{1}Q_{\beta\alpha}
&=&
Q_{\beta\alpha,\rho}\pt{1}\rho +
Q_{\beta\alpha,p_\gamma}\pt{1}p_\gamma +
Q_{\beta\alpha,\ene}\pt{1}\Ene\\
&=&
-Q_{\beta\alpha,\rho}\nabla_\gamma p_\gamma
-Q_{\beta\alpha,p_\gamma}\nabla_\delta Q_{\delta\gamma}
-Q_{\beta\alpha,\ene}\nabla_\gamma K_\gamma\\
&=&
-Q_{\beta\alpha,\rho}\nabla_\gamma p_\gamma
-Q_{\beta\alpha,p_\gamma}
\left(
 Q_{\delta\gamma,\rho}\nabla_\delta\rho +
 Q_{\delta\gamma,p_\xi}\nabla_\delta p_\xi +
 Q_{\delta\gamma,\Ene}\nabla_\delta\Ene
\right)\\
& &
-Q_{\beta\alpha,\ene}
\left(
 K_{\gamma,\rho}\nabla_\gamma\rho +
 K_{\gamma,p_\xi}\nabla_\gamma p_\xi +
 K_{\gamma,\Ene}\nabla_\gamma\Ene
\right)\\
&=&
-\left(
 Q_{\beta\alpha,p_\gamma}Q_{\delta\gamma,\rho} +
 Q_{\beta\alpha,\Ene}K_{\delta,\rho}
\right)\nabla_\delta\rho\\
& &
-\left(
 Q_{\beta\alpha,\rho}\delta_{\delta\xi} +
 Q_{\beta\alpha,p_\gamma}Q_{\delta\gamma,p_\xi} +
 Q_{\beta\alpha,\Ene}K_{\delta,p_\xi}
\right)\nabla_\delta p_\xi\\
& &
-\left(
 Q_{\beta\alpha,p_\gamma}Q_{\delta\gamma,\Ene} +
 Q_{\beta\alpha,\Ene}K_{\delta,\Ene}
\right)\nabla_\delta\Ene,
\end{eqnarray*}
and likewise
\begin{eqnarray*}
\pt{1}K_\beta
&=&
-\left(
 K_{\beta,p_\gamma}Q_{\delta\gamma,\rho} +
 K_{\beta,\Ene}K_{\delta,\rho}
\right)\nabla_\delta\rho\\
& &
-\left(
 K_{\beta,\rho}\delta_{\delta\xi} +
 K_{\beta,p_\gamma}Q_{\delta\gamma,p_\xi} +
 K_{\beta,\Ene}K_{\delta,p_\xi}
\right)\nabla_\delta p_\xi\\
& &
-\left(
 K_{\beta,p_\gamma}Q_{\delta\gamma,\Ene} +
 K_{\beta,\Ene}K_{\delta,\Ene}
\right)\nabla_\delta\Ene,
\end{eqnarray*}
where the commas denote differentiation~\footnote{Thus, for example,
  $K_{\beta,\rho}\equiv\partial K_\beta/\partial\rho$.}.  Inserting
these into Eqs.~(\ref{eq:sop0}) and (\ref{eq:soe0}), we get
\begin{eqnarray}
\pt{2}p_\alpha + \nabla_\beta Q_{\beta\alpha} &=&
 \left(\tau-\frac{1}{2}\right)
 \nabla_\beta
 \left(
  \calDc{\bfp\rho}{\beta\alpha\delta}\nabla_\delta\rho +
  \calDc{\bfp\bfp}{\beta\alpha\delta\xi}\nabla_\delta p_\xi +
  \calDc{\bfp\Ene}{\beta\alpha\delta}\nabla_\delta\Ene
 \right)\nonumber\\
\pt{2}\Ene + \nabla_\beta K_\beta &=&
 \left(\tau-\frac{1}{2}\right)
 \nabla_\beta
 \left(
  \calDc{\Ene\rho}{\beta\delta}\nabla_\delta\rho +
  \calDc{\Ene\bfp}{\beta\delta\xi}\nabla_\delta p_\xi +
  \calDc{\Ene\Ene}{\beta\delta}\nabla_\delta\Ene
 \right),
\label{eq:wyg0}
\end{eqnarray}
where we have defined the transport coefficients
\begin{eqnarray}
\calDc{\bfp\rho}{\beta\alpha\delta} &\equiv&
 R_{\delta\beta\alpha,\rho} -
 Q_{\beta\alpha,p_\gamma}Q_{\delta\gamma,\rho} -
 Q_{\beta\alpha,\Ene}K_{\delta,\rho}\nonumber\\
\calDc{\bfp\bfp}{\beta\alpha\delta\xi} &\equiv& 
 R_{\delta\beta\alpha,p_\xi} -
 Q_{\beta\alpha,\rho}\delta_{\delta\xi} -
 Q_{\beta\alpha,p_\gamma}Q_{\delta\gamma,p_\xi} -
 Q_{\beta\alpha,\Ene}K_{\delta,p_\xi}\nonumber\\
\calDc{\bfp\Ene}{\beta\alpha\delta} &\equiv& 
 R_{\delta\beta\alpha,\Ene} -
 Q_{\beta\alpha,p_\gamma}Q_{\delta\gamma,\Ene} -
 Q_{\beta\alpha,\Ene}K_{\delta,\Ene}\nonumber\\
\calDc{\Ene\rho}{\beta\delta} &\equiv& 
 L_{\delta\beta,\rho} -
 K_{\beta,p_\gamma}Q_{\delta\gamma,\rho} -
 K_{\beta,\Ene}K_{\delta,\rho}\nonumber\\
\calDc{\Ene\bfp}{\beta\delta\xi} &\equiv& 
 L_{\delta\beta,p_\xi} -
 K_{\beta,\rho}\delta_{\delta\xi} -
 K_{\beta,p_\gamma}Q_{\delta\gamma,p_\xi} -
 K_{\beta,\Ene}K_{\delta,p_\xi}\nonumber\\
\calDc{\Ene\Ene}{\beta\delta} &\equiv& 
 L_{\delta\beta,\Ene} -
 K_{\beta,p_\gamma}Q_{\delta\gamma,\Ene} -
 K_{\beta,\Ene}K_{\delta,\Ene}.\label{eq:tcoefs0}
\end{eqnarray}
These are the most general equations that we can obtain using the
lattice Boltzmann method that we have described.  Note that $\bfQ$ and
$\bfK$ are determined by the desired momentum and energy fluxes.
Eqs.~(\ref{eq:tcoefs0}) then give us differential equations for the
moments $\bfR$ and $\bfL$ in terms of the conserved densities.

\section{Desired Thermohydrodynamic Equations}
\label{ssec:des}

The equations of thermohydrodynamics can be written in conservative form
as follows
\begin{eqnarray}
0 &=&
\partial_t\rho +
\bfnabla\cdot\left(\rho\bfu\right)
\label{eq:hydm0}\\
0 &=&
\partial_t\left(\rho\bfu\right) +
\bfnabla\cdot\left(\rho\bfu\bfu - \bfP\right)
\label{eq:hydp0}\\
0 &=&
\partial_t\left(\rho\ene\right) +
\bfnabla\cdot
\left(
\rho\bfu\ene - \bfu\cdot\bfP - k\bfnabla\iota
\right)
\label{eq:hyde0}
\end{eqnarray}
where $\rho$ is the density, $\bfu$ is the hydrodynamic velocity,
$\ene$ is the total energy per unit mass, and
\[
\iota\equiv\ene-\frac{u^2}{2}
\]
is the {\it internal} energy per unit mass.  Also, we have defined
\[
\bfP\equiv
-P\bfone + \lambda\left(\bfnabla\cdot\bfu\right)\bfone +
\mu\left[\bfnabla\bfu + \left(\bfnabla\bfu\right)^T\right],
\]
where $P$ is the pressure.

Comparing the conserved densities in the above equations with those used
in the last subsection, we see that the meaning of $\rho$ is unchanged
and that we must identify
\begin{eqnarray*}
\bfp &=& \rho\bfu\\
\Ene &=& \rho\ene.
\end{eqnarray*}
We also define the internal energy density,
\[
I\equiv\rho\iota = \Ene-\frac{p^2}{2\rho},
\]
and assume henceforth that the pressure $P(\rho,I)$ is a function only
of the mass density and the internal energy density.  Recasting
Eqs.~(\ref{eq:hydm0}) through (\ref{eq:hyde0}) in terms of $\rho$,
$\bfp$ and $\Ene$ is then straightforward, and after some algebra we
find
\begin{eqnarray}
\partial_t p_\alpha +
\nabla_\beta\left(\frac{p_\beta p_\alpha}{\rho} +
P\delta_{\beta\alpha}\right)
&=&
\left(\tau-\frac{1}{2}\right)
\nabla_\beta\left[
\bar{\Lambda}_{\alpha\beta\delta\xi}\left(
\nabla_\delta p_\xi - \frac{p_\xi}{\rho}\nabla_\delta\rho\right)
\right]\nonumber\\
\partial_t\Ene +
\nabla_\beta\left[
\frac{1}{\rho}\left(\Ene + P\right) p_\beta\right]
&=&
\left(\tau-\frac{1}{2}\right)
\nabla_\beta\left[
\bar{\Lambda}_{\alpha\beta\delta\xi}\left(\nabla_\delta p_\xi -
\frac{p_\xi}{\rho}\nabla_\delta\rho\right)
\frac{p_\alpha}{\rho}
\right.\nonumber\\
& &
\left.
\phantom{\nabla}
-\frac{\bar{k}\delta_{\beta\delta}}{\rho}\left(\Ene - \frac{p^2}{\rho}\right)
\nabla_\delta\rho -
\frac{\bar{k}}{\rho}
\delta_{\beta\delta}p_\xi\nabla_\delta p_\xi +
\bar{k}\delta_{\beta\delta}\nabla_\delta\Ene
\right],
\label{eq:wys0}
\end{eqnarray}
where we have defined
\[
\bar{\Lambda}_{\alpha\beta\delta\xi}\equiv
\bar{\lambda}\delta_{\alpha\beta}\delta_{\delta\xi} +
\bar{\mu}\delta_{\beta\xi}\delta_{\delta\alpha} +
\bar{\mu}\delta_{\alpha\xi}\delta_{\delta\beta}
\]
and
\begin{eqnarray*}
\bar{\lambda} &\equiv& \frac{\lambda}{\left(\tau-1/2\right)\rho}\\
\bar{\mu} &\equiv& \frac{\mu}{\left(\tau-1/2\right)\rho}\\
\bar{k} &\equiv& \frac{k}{\left(\tau-1/2\right)\rho}.
\end{eqnarray*}
That is, we have incorporated the factor of $\tau-1/2$ into the
definitions of the barred transport coefficients to facilitate
comparison of Eqs.~(\ref{eq:wys0}) and (\ref{eq:wyg0}).  Henceforth, we
assume that $\bar{\lambda}$, $\bar{\mu}$ and $\bar{k}$ are, like the
pressure $P$, functions of the mass density $\rho$ and internal energy
density $I$ only.

We can now identify the required momentum and energy fluxes,
\begin{eqnarray}
Q_{\beta\alpha} &=& \frac{p_\beta p_\alpha}{\rho}+
 P\delta_{\beta\alpha}\nonumber\\
K_\beta &=& \frac{p_\beta}{\rho}\left(\Ene+P\right).
\label{eq:fluxes0}
\end{eqnarray}
Thus, the second moments, $Q_{\alpha\beta}$, and the trace of the third
moments, $R_{\alpha\alpha\beta}=2K_\beta$, are completely determined in
terms of the conserved densities.  To proceed, we have to show that
third and (traces of the) fourth moments with the required properties
exist.  Necessary and sufficient conditions for this are discussed in
the next subsection.

\section{Consistency Conditions}

To determine the tensors $\bfR$ and $\bfL$, we compare Eqs.~(\ref{eq:wys0})
and (\ref{eq:wyg0}) to make the identification
\begin{eqnarray}
\calDc{\bfp\rho}{\beta\alpha\delta} &=&
 -\frac{p_\xi}{\rho}\bar{\Lambda}_{\alpha\beta\delta\xi}\\
\calDc{\bfp\bfp}{\beta\alpha\delta\xi} &=& 
 \bar{\Lambda}_{\alpha\beta\delta\xi}\\
\calDc{\bfp\Ene}{\beta\alpha\delta} &=& 
 0\\
\calDc{\Ene\rho}{\beta\delta} &=& 
 -\frac{p_\alpha p_\xi}{\rho^2}\bar{\Lambda}_{\alpha\beta\delta\xi}-
 \frac{\bar{k}\delta_{\beta\delta}}{\rho}\left(\Ene-\frac{p^2}{\rho}\right)\\
\calDc{\Ene\bfp}{\beta\delta\xi} &=& 
 \frac{p_\alpha}{\rho}\bar{\Lambda}_{\alpha\beta\delta\xi}-
 \frac{\bar{k}p_\xi}{\rho}\delta_{\beta\delta}\\
\calDc{\Ene\Ene}{\beta\delta} &=& 
 \bar{k}\delta_{\beta\delta}.
\end{eqnarray}
Hence, from Eq.~(\ref{eq:tcoefs0}) we have
\begin{eqnarray}
R_{\delta\beta\alpha,\rho} &\equiv&
 Q_{\beta\alpha,p_\gamma}Q_{\delta\gamma,\rho} +
 Q_{\beta\alpha,\Ene}K_{\delta,\rho} -
 \frac{p_\xi}{\rho}\bar{\Lambda}_{\alpha\beta\delta\xi}\nonumber\\
R_{\delta\beta\alpha,p_\xi} &\equiv& 
 Q_{\beta\alpha,\rho}\delta_{\delta\xi} +
 Q_{\beta\alpha,p_\gamma}Q_{\delta\gamma,p_\xi} +
 Q_{\beta\alpha,\Ene}K_{\delta,p_\xi} +
 \bar{\Lambda}_{\alpha\beta\delta\xi}\nonumber\\
R_{\delta\beta\alpha,\Ene} &\equiv& 
 Q_{\beta\alpha,p_\gamma}Q_{\delta\gamma,\Ene} +
 Q_{\beta\alpha,\Ene}K_{\delta,\Ene},
\label{eq:rvec0}
\end{eqnarray}
and
\begin{eqnarray}
L_{\delta\beta,\rho} &\equiv&
 K_{\beta,p_\gamma}Q_{\delta\gamma,\rho} +
 K_{\beta,\Ene}K_{\delta,\rho} -
 \frac{p_\alpha p_\xi}{\rho^2}\bar{\Lambda}_{\alpha\beta\delta\xi}-
 \frac{\bar{k}\delta_{\beta\delta}}{\rho}\left(\Ene-\frac{p^2}{\rho}\right)\nonumber\\
L_{\delta\beta,p_\xi} &\equiv& 
 K_{\beta,\rho}\delta_{\delta\xi} +
 K_{\beta,p_\gamma}Q_{\delta\gamma,p_\xi} +
 K_{\beta,\Ene}K_{\delta,p_\xi} +
 \frac{p_\alpha}{\rho}\bar{\Lambda}_{\alpha\beta\delta\xi}-
 \frac{\bar{k}p_\xi}{\rho}\delta_{\beta\delta}\nonumber\\
L_{\delta\beta,\Ene} &\equiv& 
 K_{\beta,p_\gamma}Q_{\delta\gamma,\Ene} +
 K_{\beta,\Ene}K_{\delta,\Ene} +
 \bar{k}\delta_{\beta\delta}.
\label{eq:lvec0}
\end{eqnarray}
Since Eqs.~(\ref{eq:fluxes0}) give the components of $\bfQ$ and $\bfK$
in terms of conserved densities, it follows that the entire right-hand
sides of Eqs.~(\ref{eq:rvec0}) and (\ref{eq:lvec0}) are known in terms
of the conserved densities.  Eqs.~(\ref{eq:rvec0}) and (\ref{eq:lvec0})
then give the partial derivatives of $\bfR$ and $\bfL$, respectively,
with respect to the three conserved densities.  Thus, analyticity of
$\bfR$ and $\bfL$ requires that all mixed second derivatives be
symmetric; in addition, we must demand that the trace of $\bfR$ be equal
to $2\bfK$, as required.  The determination of the second derivatives is
straightforward in principle but very tedious, being somewhat simplified
by the assumption that $P$, $\bar{\lambda}$, $\bar{\mu}$ and $\bar{k}$
depend only on $\rho$ and $I$.  The results are presented in
Appendix~\ref{sec:app}.  A glance at this appendix shows that, even
accounting for the isotropy and symmetry under index interchange of
$R_{\delta\beta\alpha}$ and $L_{\delta\beta}$ and of the second
derivatives with respect to the momentum density, there are dozens of
potentially independent requirements to be satisfied.  It is therefore
something of a minor miracle that all of these requirements can be
reduced to the following six scalar conditions
\begin{eqnarray}
0 &=& \bar{\mu}_{,\rho}+\frac{\bar{\mu}-P_{,\rho}}{\rho}
\label{eq:mud}\\
0 &=& \bar{\mu}_{,I}-\frac{P_{,I}}{\rho}
\label{eq:mui}\\
0 &=& \bar{\lambda}_{,\rho} + \frac{\bar{\lambda}+P_{,\rho}P_{,I}}{\rho} +
      \left(\frac{I+P}{\rho}\right)P_{,\rho,I} + P_{,\rho,\rho}
\label{eq:lmd}\\
0 &=& \bar{\lambda}_{,I} + \frac{P_{,I}^2}{\rho} +
      \left(\frac{I+P}{\rho}\right)P_{,I,I} + P_{,\rho,I}
\label{eq:lmi}\\
0 &=& \bar{\lambda}-\bar{\mu} + \left(\frac{I+P}{\rho}\right)P_{,I} +
      P_{,\rho}
\label{eq:eosres}\\
0 &=& \bar{\lambda}-\bar{\mu}+\bar{k}+I\bar{k}_{,I}+\rho\bar{k}_{,\rho}
\label{eq:kpd}
\end{eqnarray}
in {\it any} number of dimensions.  The first pair of these,
Eqs.~(\ref{eq:mud}) and (\ref{eq:mui}), must be integrated for
$\bar{\mu}$; analyticity then imposes a consistency condition on the
mixed second derivatives of $\bar{\mu}$, but it is seen to be satisfied
identically, and these equations then integrate to yield
\begin{equation}
\bar{\mu} = \frac{P}{\rho}.
\label{eq:muans}
\end{equation}
Likewise, the second pair, Eqs.~(\ref{eq:lmd}) and (\ref{eq:lmi}) for
$\bar{\lambda}$, are automatically consistent with analytic
$\bar{\lambda}$, and integrate to yield
\begin{equation}
\bar{\lambda} = \frac{P}{\rho}-P_{,\rho}-
\left(\frac{I+P}{\rho}\right)P_{,I}.
\label{eq:lamans}
\end{equation}
The fifth equation, Eq.~(\ref{eq:eosres}), is then satisfied
identically.  This leaves Eq.~(\ref{eq:kpd}) which is a partial
differential equation for the thermal conductivity; we shall return to
this equation in a moment.

With the aid of Eqs.~(\ref{eq:muans}) and (\ref{eq:lamans}),
Eqs.~(\ref{eq:rvec0}) can be integrated to get
\[
R_{\delta\beta\alpha} =
\left(
\delta_{\beta\alpha}p_\delta +
\delta_{\delta\alpha}p_\beta +
\delta_{\delta\beta}p_\alpha
\right) \frac{P}{\rho} +
\frac{p_\delta p_\beta p_\alpha}{\rho^2}.
\]
For $D=3$, this result may be compared to Equation (8d) in the reference
by McNamara~\cite{bib:mac}.  We must now demand that the trace of this
quantity be equal to twice the heat flux.  We find
\[
R_{\delta\beta\beta}=
\left[(D+2)\frac{P}{\rho}+\frac{p^2}{\rho^2}\right] p_\delta,
\]
where $D=\delta_{\alpha\alpha}$ is the spatial dimension.  Equating this
with twice the heat flux, as given by Eq.~(\ref{eq:fluxes0}), we arrive at
\[
P=\frac{2}{D}I,
\]
which is the ideal gas law~\footnote{To see this in a more familiar
  form, note that the internal energy per unit mass is
  $I=\smallfrac{D}{2}\rho k_B T/m$, so that $P=n k_B T$ where $n=\rho/m$
  is the number density.}.  From this, Eq.~(\ref{eq:muans}) immediately
  yields
\[
\bar{\mu} = \frac{2I}{D\rho},
\]
and Eq.~(\ref{eq:lamans}) yields
\[
\bar{\lambda} = -\frac{4I}{D^2\rho}.
\]

Returning to the thermal conductivity, it is straightforward to
establish, e.g. using the method of characteristics, that
Eq.~(\ref{eq:kpd}) has the general solution
\[
\bar{k} = \frac{2(D+2)}{D^2\rho}
\left[I+f'\left(\frac{I}{\rho}\right)\right],
\]
where $f'$ is (the derivative of) an arbitrary function of its argument.
Thus, we find that the model does have some flexibility in the choice of
thermal conductivity.  Using this result, it is now straightforward to
integrate Eqs.~(\ref{eq:lvec0}) for the trace of the fourth moment, and
the result is
\[
L_{\delta\beta}
=
\frac{2(D+2)}{D\rho}
\left\{
\frac{1}{D}
\left[
\left(
I+\frac{Dp^2}{2(D+2)\rho}
\right)I +
\rho f\left(\frac{I}{\rho}\right)
\right]\delta_{\delta\beta}
+\frac{D+4}{D+2}
\left(
I+\frac{Dp^2}{2(D+4)\rho}
\right) \frac{p_\delta p_\beta}{\rho}
\right\}.
\]
For $D=3$ and $f=0$ this may be compared to Equation (8c) in the paper
by McNamara~\cite{bib:mac}.  To the best of our knowledge, however, this
form for $\bfL$ has never been written down for general $D$, or with the
$f$ term that allows some control of the thermal conductivity.

The ``unbarred'' forms of the transport coefficients are then,
\begin{eqnarray*}
\mu &=& \frac{2}{D}\left(\tau-\smallfrac{1}{2}\right)I\\
\lambda &=& -\frac{4}{D^2}\left(\tau-\smallfrac{1}{2}\right)I\\
k &=& \frac{2(D+2)}{D^2}\left(\tau-\smallfrac{1}{2}\right)
\left[I+f'\left(\frac{I}{\rho}\right)\right].
\end{eqnarray*}
We note that $\mu$ and $\lambda$ are independent of $\rho$, while $k$
depends on it only via the arbitrary function $f$.  We see that the
relation
\[
\lambda=-\frac{2}{D}\mu
\]
holds; when $D=3$ this is approximately true for many real gases.  The
linear temperature dependence of $\lambda$ and $\mu$ are, however,
unrealistic; for real gases these go as $\sqrt{I}$.  Finally, the
Prandtl number is
\[
\mbox{Pr} \equiv \frac{\mu}{k} =
\frac{D}{D+2}
\left[1 + \frac{f'(I/\rho)}{I}\right]^{-1}.
\]
For $f=0$, this reduces to the constant $D/(D+2)$ which is indeed equal
to $1/2$ for the $D=2$ example treated by Alexander et
al.~\cite{bib:alexander}, and $3/5$ for the $D=3$ example treated by
McNamara~\cite{bib:mac} (when the relaxation times in the latter paper
are set equal).  More generally, however, the function $f$ may be used
to exercise some control over the Prandtl number.

\section{Conclusions}

We have presented a first-principles derivation of the lattice-BGK model
of the ideal gas.  In the process, we have established the necessity of
the functional forms used for the shear and bulk viscosities, and have
derived a somewhat more general form than is usually assumed for the
thermal conductivity.  We have also shown that the thermohydrodynamic
model of the ideal gas is grossly overspecified, so that its existence
is largely fortuitous.  We have also shown that the unphysical
dependence of the viscosities on temperature is an essential feature of
the lattice-BGK model with constant relaxation time.

The cornerstone of our demonstration was nothing more than the demand of
analyticity of the third moment and the trace of the fourth moment of
the equilibrium distribution function.  The main point of this paper is
that it is essential to pay close attention to this criterion when using
the ``inverse Chapman-Enskog'' method of derivation of lattice Boltzmann
models.  It is a very fundamental and restrictive requirement which
cannot be circumvented by different choices of lattice or of velocity
set, or other superficial details.  As noted earlier, allowing the
relaxation parameter to depend on the hydrodynamic densities may be a
way to recover some flexibility in this regard, but this possibility
remains completely unexplored.

One of the principal outstanding problems in lattice Boltzmann research
is the development of a thermohydrodynamically consistent model that
includes a soft interaction potential~\cite{bib:yeo,bib:shch}.  The
ultimate solution to this problem will surely involve a radical
departure from current lattice Boltzmann models -- for example, by
inclusion of some information about the two-particle distribution
function.  However it is solved, we expect that the form of the
equilibrium distribution function involved will not be as easy to intuit
as it is for the ideal gas.  In that event, we expect that an approach
similar to the one that was adpoted in this paper will be useful.

\section*{Acknowledgements}

The authors would like to thank Frank Alexander and Julia Yeomans for
helpful discussions, and Schlumberger Cambridge Research and NATO grant
number CRG950356 for supporting this collaboration.  BMB was supported
in part by the United States Air Force Office of Scientific Research
under grant number F49620-95-1-0285, and in part by AFRL.

\newpage
\appendix

\section{Mixed Second Derivatives of the Moments}
\label{sec:app}

The mixed second derivatives of the third moments and the trace of the
fourth moments of the equilibrium distribution function can be
calculated as described in Secton 7.  The results are
\begin{eqnarray*}
R_{\delta\beta\alpha,\rho,p_\xi} -
R_{\delta\beta\alpha,p_\xi,\rho}
&=&
-\left(
\delta_{\delta\beta}\delta_{\alpha\xi} + \delta_{\delta\alpha}\delta_{\beta\xi}
\right)
\left(
\bar{\mu}_{,\rho} + \frac{\bar{\mu}-P_{,\rho}}{\rho}
\right)\\
& &
-\delta_{\delta\xi}\delta_{\beta\alpha}
\left[
\bar{\lambda}_{,\rho} + \frac{\bar{\lambda}+P_{,\rho}P_{,I}}{\rho} +
\left(\frac{I+P}{\rho}\right)P_{,\rho,I} + P_{,\rho,\rho}
\right]\\
& &
+\frac{1}{\rho^2}
\left\{
\left[
\delta_{\delta\beta}\left(p_\alpha p_\xi - \frac{p^2}{2}\delta_{\alpha\xi}\right) +
\delta_{\delta\alpha}\left(p_\beta p_\xi - \frac{p^2}{2}\delta_{\beta\xi}\right)
\right]
\left(
\bar{\mu}_{,I}-\frac{P_{,I}}{\rho}
\right)\right.\\
& &
\left. +
\left[
\delta_{\beta\alpha}\left(p_\delta p_\xi - \frac{p^2}{2}\delta_{\delta\xi}\right)
\right]
\left[
\bar{\lambda}_{,I} + \frac{P_{,I}^2}{\rho} +
\left(\frac{I+P}{\rho}\right)P_{,I,I} + P_{,\rho,I}
\right]\right\}\\
R_{\delta\beta\alpha,\rho,\Ene} -
R_{\delta\beta\alpha,\Ene,\rho}
&=&
-\frac{p_\delta}{\rho}
\delta_{\beta\alpha}
\left[
\bar{\lambda}_{,I} +
\frac{P_{,I}^2}{\rho} +
\left(\frac{I+P}{\rho}\right) P_{,I,I} +
P_{,I,\rho}
\right]\\
& &
-\frac{1}{\rho}
\left(
 \delta_{\alpha\delta}p_\beta +
 \delta_{\beta\delta}p_\alpha
\right)
\left(
 \bar{\mu}_{,I} - \frac{P_{,I}}{\rho}
\right)\\
R_{\delta\beta\alpha,p_\xi,p_\eta} -
R_{\delta\beta\alpha,p_\eta,p_\xi}
&=&
+\frac{1}{\rho}
\left[
\left(
\delta_{\delta\beta}\delta_{\alpha\eta} +
\delta_{\beta\eta}\delta_{\delta\alpha}
\right) p_\xi -
\left(
\delta_{\delta\alpha}\delta_{\beta\xi} +
\delta_{\delta\beta}\delta_{\alpha\xi}
\right) p_\eta
\right]
\left(
\bar{\mu}_{,I} - \frac{P_{,I}}{\rho}
\right)\\
& &
+\frac{1}{\rho}
\left(
\delta_{\delta\eta}\delta_{\beta\alpha} p_\xi -
\delta_{\delta\xi}\delta_{\beta\alpha} p_\eta
\right)
\left[
\bar{\lambda}_{,I} + \frac{P_{,I}^2}{\rho} +
\left(\frac{I+P}{\rho}\right) P_{,I,I} +
P_{,\rho,I}
\right]\\
R_{\delta\beta\alpha,p_\xi,\Ene} -
R_{\delta\beta\alpha,\Ene,p_\xi}
&=&
+\delta_{\delta\xi}\delta_{\beta\alpha}
\left[
\bar{\lambda}_{,I} +
\frac{P_{,I}^2}{\rho} +
\left(\frac{I+P}{\rho}\right) P_{,I,I} +
P_{,I,\rho}
\right]\\
& &
+\left(
 \delta_{\beta\xi}\delta_{\alpha\delta} +
 \delta_{\alpha\xi}\delta_{\beta\delta}
\right)
\left(
 \bar{\mu}_{,I} - \frac{P_{,I}}{\rho}
\right)\\
L_{\delta\beta,\rho,p_\xi} -
L_{\delta\beta,p_\xi,\rho}
&=&
+\frac{\delta_{\delta\beta}p_\xi}{\rho^2}
\left\{
\left[
\bar{k}+I\bar{k}_{,I}+\rho\bar{k}_{,\rho}-\mu-\rho\mu_{,\rho}-
\left(\frac{I+P}{\rho}\right)P_{,I}
\right]
\right.\\
& &
\left.
+\frac{p^2}{2\rho}
\left(
\bar{\mu}_{,I}-\frac{P_{,I}}{\rho}
\right)
\right\}\\
& &
+\frac{\delta_{\delta\xi}p_\beta}{\rho^2}
\left\{
\left[
\bar{\mu}-\left(\frac{I+P}{\rho}\right)\left(P_{,I}+\rho P_{,\rho,I}\right)+
\rho\bar{\lambda}_{,\rho}-P_{,\rho}
\right.
\right.\\
& &
\left.
\left.
+P_{,I}P_{,\rho}+\rho P_{,\rho,\rho}
\right]
-\frac{p^2}{2\rho}
\left[
\bar{\lambda}_{,I}+\frac{P_{,I}^2}{\rho}+
\left(\frac{I+P}{\rho}\right) P_{,I,I}+P_{,\rho,I}
\right]
\right\}\\
& &
-\frac{\delta_{\beta\xi}p_\delta}{\rho^2}
\left\{
\left[
\bar{\lambda}+\left(\frac{I+P}{\rho}\right)P_{,I}+\rho\bar{\mu}_{,\rho}
\right]
+\frac{p^2}{2\rho}
\left(
\bar{\mu}_{,I}-\frac{P_{,I}}{\rho}
\right)
\right\}\\
& &
-\rho^2p_ip_jp_k
\left[
\bar{\lambda}_{,I}+
\frac{P_{,I}^2}{\rho}+
\left(\frac{I+P}{\rho}\right)P_{,I,I}+
P_{,\rho,I}+
\bar{\mu}_{,I}-
\frac{P_{,I}}{\rho}
\right]\\
L_{\delta\beta,\rho,\Ene} -
L_{\delta\beta,\Ene,\rho}
&=&
-\frac{\delta_{\delta\beta}}{\rho}
\left\{
\left[
\bar{k}+I\bar{k}_{,I}+\rho\bar{k}_{,\rho}-
\left(\frac{I+P}{\rho}\right)P_{,I}-P_{,\rho}
\right]
+\frac{p^2}{\rho}
\left(
\bar{\mu}_{,I}-\frac{P_{,I}}{\rho}
\right)
\right\}\\
& &
-\frac{p_ip_j}{\rho^2}
\left[
\bar{\lambda}_{,I}+
\frac{P_{,I}^2}{\rho}+
\left(\frac{I+P}{\rho}\right)P_{,I,I}+
P_{,\rho,I}+
\bar{\mu}_{,I}-
\frac{P_{,I}}{\rho}
\right]\\
L_{\delta\beta,p_\xi,p_\eta} -
L_{\delta\beta,p_\eta,p_\xi}
&=&
\frac{\delta_{\delta\xi}\delta_{\beta\eta} -
      \delta_{\delta\eta}\delta_{\beta\xi}}{\rho}
\left[
\bar{\lambda}-\bar{\mu}+
\left(\frac{I+P}{\rho}\right)P_{,I}+P_{,\rho}
\right]\\
& &
+\frac{
p_\delta
\left(
\delta_{\beta\eta}p_\xi - \delta_{\beta\xi}p_\eta
\right)}{\rho^2}
\left(
\bar{\mu}_{,I}-\frac{P_{,I}}{\rho}
\right)\\
& &
+\frac{
p_\beta
\left(
\delta_{\delta\eta}p_\xi - \delta_{\delta\xi}p_\eta
\right)}{\rho^2}
\left[
\bar{\lambda}_{,I}+\frac{P_{,I}^2}{\rho}+
\left(\frac{I+P}{\rho}\right)P_{,I,I}+P_{,\rho,I}
\right]\\
L_{\delta\beta,p_\xi,\Ene} -
L_{\delta\beta,\Ene,p_\xi}
&=&
+\frac{1}{\rho}
\left(
\delta_{\delta\beta}p_\xi+\delta_{\beta\xi}p_\delta
\right)
\left(
\bar{\mu}_{,I}-\frac{P_{,I}}{\rho}
\right)\\
& &
+\frac{\delta_{\delta\xi}p_\beta}{\rho}
\left[
\bar{\lambda}_{,I}+\frac{P_{,I}^2}{\rho}+
\left(\frac{I+P}{\rho}\right)P_{,I,I}+P_{,\rho,I}
\right].
\end{eqnarray*}

\end{document}